\documentclass[12pt,twoside,a4paper]{article}
\pdfoutput=1
\usepackage{graphicx}
\usepackage{url}
\usepackage{color}
\usepackage{cite}

\def\lsim{\:\raisebox{-0.5ex}{$\stackrel{\textstyle<}{\sim}$}\:}
\def\gsim{\:\raisebox{-0.5ex}{$\stackrel{\textstyle>}{\sim}$}\:}
\begin{document}
\thispagestyle{empty} 
\title{
\vskip-3cm
{\baselineskip14pt
\centerline{\normalsize DESY 19-115 \hfill ISSN 0418--9833}
\centerline{\normalsize MITP/19--045 \hfill} 
\centerline{\normalsize July 2019 \hfill}} 
\vskip1.5cm
\boldmath
{\bf $B$-meson production in the}
\\
{\bf general-mass variable-flavour-number scheme} 
\\
{\bf and LHC data}
\unboldmath
\author{
M. Benzke$^1$,
B. A. Kniehl$^2$\footnote{On leave of absence from 
  II. Institut f\"ur Theoretische
  Physik, Universit\"at Hamburg, Luruper Chaussee 149, 
  D-22761 Hamburg, Germany} , 
G. Kramer$^1$, 
I. Schienbein$^3$ 
and H. Spiesberger$^4$
\vspace{2mm} \\
\normalsize{
  $^1$ II. Institut f\"ur Theoretische
  Physik, Universit\"at Hamburg,
}\\ 
\normalsize{
  Luruper Chaussee 149, D-22761 Hamburg, Germany
} \vspace{2mm}\\
\normalsize{
  $^2$ University of California at San Diego, 9500 Gilman Drive, 
  La Jolla, CA 92093, USA
} \vspace{2mm}\\
\normalsize{
  $^3$ Laboratoire de Physique Subatomique et de Cosmologie,
} \\ 
\normalsize{
  Universit\'e Joseph Fourier Grenoble 1,
}\\
\normalsize{
  CNRS/IN2P3, Institut National Polytechnique de Grenoble,
}\\
\normalsize{
  53 avenue des Martyrs, F-38026 Grenoble, France
} \vspace{2mm}\\
\normalsize{
  $^4$ PRISMA{\color{red}$^{+}$} Cluster of Excellence, 
  Institut f\"ur Physik,
  Johannes-Gutenberg-Universit\"at,
}\\ 
\normalsize{
  Staudinger Weg 7, D-55099 Mainz, Germany,}\\
\normalsize{and Centre for Theoretical and
  Mathematical Physics and Department of Physics,}\\
\normalsize{
  University of Cape Town, Rondebosch 7700, South Africa}\\
\vspace{2mm} \\}}
\date{\today}
\maketitle
\begin{abstract}
\medskip
\noindent
We study inclusive $B$-meson production in $pp$ collisions 
at the LHC and compare experimental data with predictions 
of the general-mass variable-flavour-number scheme at 
next-to-leading order of perturbative QCD. We find almost 
perfect agreement provided that the factorization scale 
parameters and the parton distribution functions are chosen 
appropriately. 
\end{abstract}
\clearpage

\section{Introduction}

The study of the inclusive production of hadrons containing $b$ 
quarks, as for example $B^{\pm}$, $B^0$, $\bar{B^0}$, $B_s^0$, 
$\bar{B}_s^0$ mesons and $\Lambda_b^0$ baryons, is particularly 
important to test quantum chromodynamics (QCD). The predictions 
in the framework of perturbative QCD are based on the 
factorization approach. In this approach, the production
cross section is calculated as a convolution of three basic 
ingredients: the parton distribution functions (PDFs) describing 
the parton content of the hadronic initial state, the partonic 
hard-scattering cross section computed as a perturbative series 
in powers of the strong-coupling constant, and the fragmentation 
functions (FFs), which describe the momentum distribution for 
specified $b$ hadrons in a parton. Since the $b$-quark mass is 
large and cannot be neglected in the small transverse momentum 
region, the cross section for $b$-hadron production depends on 
several large scales, which makes predictions of the cross 
sections very demanding.

In the past, measurements of inclusive $b$-hadron production 
and the corresponding perturbative QCD calculations have been 
performed for all of the $B$ mesons mentioned above, and for 
$\Lambda_b^0$ baryons. Some time ago, data for $p\bar{p}$ 
collisions at center-of-mass energy $\sqrt{S} = 1.96$~TeV have 
been obtained at the FNAL Tevatron Collider by the CDF 
Collaboration \cite{Acosta:2004yw,Abulencia:2006ps} and later 
for $pp$ collisions at $\sqrt{S} = 5$, 7, 8 and 13~TeV at the 
CERN Large Hadron Collider (LHC) by the ATLAS, CMS and LHCb 
Collaborations 
\cite{Sirunyan:2017oug,ATLAS:2013cia, Khachatryan:2011mk, 
Chatrchyan:2011pw,Chatrchyan:2011vh,Aaij:2012jd, Aaij:2015rla, 
Khachatryan:2016csy}. First measurements of the inclusive 
production cross sections of $\Lambda_b^0$ baryons have been 
presented by the CMS Collaboration \cite{Chatrchyan:2012xg} 
for $\sqrt{S} = 7$~TeV and by the LHCb Collaboration for 
$\sqrt{S} = 7$ and 8~TeV \cite{Aaij:2015fea}.

Almost all these data have been compared with 
next-to-leading-order (NLO) QCD predictions based on the 
so-called fixed-order next-to-leading-logarithmic (FONLL) 
approach \cite{Cacciari:2012ny}. Data of the CMS, LHCb and ATLAS 
Collaborations have also been compared with predictions 
obtained in the general-mass variable-flavour-number scheme 
(GM-VFNS) \cite{Kniehl:2011bk,Kniehl:2015fla}. The GM-VFNS 
as originally formulated in Refs.~\cite{Kniehl:2004fy,Kniehl:2005mk} 
(see also Ref.~\cite{Helenius:2018uul} for a more recent
implementation of the GM-VFNS including its application to 
charm meson production) is similar to the FONLL scheme, but 
incorporates different assumptions concerning fragmentation 
functions, and the transition to the fixed-flavour-number 
scheme (FFNS) in the low transverse momentum region is treated 
in a different way. All comparisons between experimental data 
and theoretical predictions, both in the FONLL and GM-VFNS 
approaches, have shown reasonable agreement within experimental 
errors and the theoretical uncertainty, which is usually 
estimated by a variation of the factorization and 
renormalization scale parameters and the bottom quark mass.

For large transverse momenta $p_T$, say $p_T \gsim 8$~GeV, the 
two approaches yield approximately the same differential cross 
sections $d\sigma/dp_T$ as a function of $p_T$. Most of the 
data from the CMS and ATLAS Collaborations lie in this $p_T$ 
region. Comparisons with GM-VFNS predictions can be found in 
Refs.~\cite{Kniehl:2011bk,Kniehl:2015fla}. Data in the low 
$p_T$ region, i.e.\ for $0 < p_T < 8 $~GeV, have been obtained 
by the LHCb Collaboration~\cite{Aaij:2013noa,Aaij:2017qml} and 
also by the CDF Collaboration~\cite{Acosta:2004yw,Abulencia:2006ps}, 
although at different $\sqrt{S}$ values. Tevatron 
data exist at $\sqrt{S} = 1.96$~TeV, while data from the LHCb 
Collaboration are obtained at $\sqrt{S} = 7$~TeV 
\cite{Aaij:2013noa,Aaij:2017qml} and at $\sqrt{S} = 13$~TeV 
\cite{Aaij:2017qml}. Also the rapidity ($y$) range 
in these experiments is different: $|y| \leq 1$ for the CDF 
measurements \cite{Acosta:2004yw,Abulencia:2006ps} and 
$2.0 < y < 4.5$ for the two LHCb measurements 
\cite{Aaij:2013noa,Aaij:2017qml}. The behaviour of 
$d\sigma/dp_T$ as a function of $p_T$ in the low and 
the high $p_T$ ranges is quite different. At large $p_T$, 
$d\sigma/dp_T$ is monotonically decreasing in accordance 
with the expected behaviour originating from the interplay 
of the momentum dependence of the hard-scattering cross 
section and the scale dependence of the PDFs and FFs. In 
the low $p_T$ region, $0 < p_T < 8$ GeV, $d\sigma/dp_T$ 
behaves quite differently, both for the $p\bar{p}$ data 
from CDF~\cite{Acosta:2004yw,Abulencia:2006ps} and the 
$pp$ data from LHCb~\cite{Aaij:2013noa,Aaij:2017qml}. Towards 
low $p_T$, the heavy-quark production threshold takes over 
and $d\sigma/dp_T$ reaches a maximum at $p_T \simeq 2.5$~GeV 
and then decreases towards $p_T = 0$. 

This behaviour is very well predicted theoretically within 
the FFNS as shown in previous work~\cite{Kniehl:2015fla}. 
There we have also shown how the FFNS results at small $p_T$ 
can be incorporated in the GM-VFNS by choosing appropriate 
factorization scales. We use the notation $\mu_i$ for the 
initial-state factorization scale entering the PDFs and 
$\mu_f$ for the final-state factorization scale entering 
the FFs. In Ref.~\cite{Kniehl:2015fla} these two scales were 
fixed at the same value, $\mu_i = \mu_f = \mu_F := 0.5\, 
\sqrt{p_T^2 + m_b^2}$ with $m_b = 4.5$~GeV. This leads to a 
suppression of the contribution of the $b$-quark PDF in the 
proton for $p_T \simeq 8$~GeV due to the threshold 
$\mu_{\rm thr} = m_b$ as implemented in both the PDFs and 
FFs chosen in Ref.~\cite{Kniehl:2015fla}, and we could describe 
the CDF data \cite{Acosta:2004yw,Abulencia:2006ps} very well 
in the $p_T$ range $0 < p_T < 25$ GeV and the LHCb data fairly 
well in the $p_T$ range $0 < p_T < 12$~GeV showing a maximum 
of $d\sigma/dp_T$ near $p_T \simeq 2.5$ GeV. 

In Ref.~\cite{Kniehl:2015fla}, we used factorization scales in 
such a way that the transition to the FFNS occurs at rather 
large $p_T$ values. This choice was satisfactory: the predictions 
of the GM-VFNS were found to be consistent with the data 
at $\sqrt{S} = 7$ TeV inside the theoretical error estimated 
by varying the renormalization scale by a factor of two. In the 
present work, we attempt to improve the agreement between 
LHCb data and predictions, in particular in the low $p_T$ 
region, by using a more general ansatz for the factorization 
scale, which produces the transition to the FFNS at much smaller 
$p_T$ values. This new ansatz will be described in the next section.
It was used already in the calculation of charm-meson production 
in the same kinematical region as for $B^{\pm}$-meson production 
in the present work, and it was successfully compared with the 
relevant LHCb \cite{Benzke:2017yjn}, ALICE and CDF data 
\cite{Benzke:2018sns}.

In the present work, we shall compare with the more recent
LHCb measurements at $\sqrt{S} = 7$ and $13$ TeV 
\cite{Aaij:2017qml}. These measurements extend all the way up 
to $p_T = 40$~GeV. The cross section data in five 
rapidity bins in the forward region $2.0 < y < 4.5$ are 
much more accurate than the previous data at $\sqrt{S} = 
7$~TeV \cite{Aaij:2013noa}. We hope that a comparison of 
predictions for cross sections and cross section ratios 
with these 7 and 13~TeV data will allow us to obtain 
additional information on the proton PDFs at small $x$ values.
 
The outline of our work is as follows. In Sect.\ 2, we describe 
the setup of our calculation aiming at a comparison with the 
recent LHCb $B^+$-production data and discuss in particular 
the possible choices of PDFs. In Sect.\ 3, we present numerical 
results and a comparison with the LHCb data \cite{Aaij:2017qml}. 
We start with the cross sections $d^2\sigma/dp_Tdy$ for 
$\sqrt{S} = 7$ and 13~GeV for the five rapidity bins in the 
range $2.0 < y < 4.5$ as a function of $p_T$ in the range 
between 0 and 40~GeV. Then we study the single-differential 
cross section $d\sigma/dp_T$ as a function of $p_T$ in the 
same $p_T$ range, but summed over the five $y$ bins, i.e.\ 
for $2.0 < y < 4.5$, and calculate from them the ratio of the 
cross sections for $\sqrt{S} = 13$ and 7~TeV. Finally, we also 
present the rapidity dependence of the cross section,  
$d\sigma/dy$ integrated over $p_T$ in the range $0 < p_T < 
40$~GeV and consider the 13~TeV to 7~GeV ratio. In Sect.\ 4, 
we present a summary and some concluding remarks.


\section{Setup and input}

The theoretical description of the GM-VFNS approach as well 
as various technical details of its implementation have been 
presented in our previous papers \cite{Kniehl:2004fy, 
Kniehl:2005mk}. Here we describe only the input for the 
numerical evaluation discussed below, in particular our 
choice of the proton PDF.

As a default, we use the PDF set CT14 \cite{Dulat:2015mca} at NLO. 
Other choices are ($i$) HERAPDF2.0 \cite{Abramowicz:2015mha}, 
($ii$) MMHT \cite{Harland-Lang:2014zoa}, ($iii$) NNPDF3.1 
\cite{Ball:2017nwa}. All these PDF sets are NLO parametrizations; 
the last two of them are obtained from global fits to essentially 
the same experimental data as CT14, while HERAPDF2.0 is based on 
cross section data from deep inelastic scattering at HERA only. 
All PDF parametrizations have been taken from their 
implementation in the LHAPDF library \cite{Buckley:2014ana}. 

To describe the transition of $b$ quarks to $B^+$ mesons, we 
need nonperturbative FFs. We use the $B$-meson FFs constructed 
in Ref.~\cite{Kniehl:2008zza}. They are evolved at NLO, and all 
components for the transition from gluons and light quarks to 
a $B$ meson are generated through the 
Dokshitzer-Gribov-Lipatov-Altarelli-Parisi evolution. They 
were obtained by fitting experimental data for inclusive 
$b$-hadron production in $e^+e^-$ annihilation taken by the 
ALEPH~\cite{Heister:2001jg} and OPAL~\cite{Abbiendi:2002vt} 
Collaborations at CERN LEP1 and by the SLD Collaboration 
\cite{Abe:1999ki,Abe:2002iq} at SLAC SLC. These data were all 
taken at the $Z$-boson resonance. Therefore the strong coupling 
$\alpha_s^{(n_f)}(\mu_R)$ was calculated with $n_f = 5$ active 
quark flavors and the renormalization and factorization scales 
were fixed at $\mu_R = \mu_F = m_Z$. The starting scale was 
chosen to be $\mu_0 = m_b = 4.5$~GeV. Below $\mu_F = \mu_0$, 
the light-quark (including charm) FFs and the gluon FF were 
assumed to vanish. A simple power ansatz for the $b$-quark FF 
at $\mu_0$ gave the best fit to the experimental data. Recently 
also FFs at NNLO have become available \cite{Salajegheh:2019ach}. 
They were obtained from fits which also include $b$-hadron 
production data from the DELPHI Collaboration~\cite{DELPHI:2011aa}. 

The essential ingredient in the numerical calculations 
presented below is the choice of the factorization 
scales $\mu_F = \mu_i = \mu_f$. They are fixed by 
\begin{equation}
\mu_F = 0.49 \sqrt{p_T^2 + 4m_b^2}
\label{eq:muf}
\end{equation}
with the heavy-quark mass $m_b = 4.5$~GeV. This value of $m_b$ 
equals the threshold of the $b \to B^+$ FF. The scale choice 
in Eq.\ (\ref{eq:muf}) is similar to the one used in 
Refs.~\cite{Benzke:2017yjn,Benzke:2018sns}, where a successful 
interpretation of LHCb, ALICE and CDF data was achieved. The 
factor 4 in front of $m_b^2$ can also be justified by the fact 
that the dominant LO contribution in the FFNS originates from 
the channel $gg \to b\bar{b}$ which has the threshold $2m_b$. 
Both for PDFs and FFs we keep $\mu_i$ and $\mu_f$ constant below 
the threshold value $\mu_0 = m_b$, i.e.\ the PDFs and FFs are 
not evolved when $p_T$ is further decreased and $\mu_F$, as a 
function of $p_T$, becomes smaller than $m_b$. With the 
definition in Eq.~(\ref{eq:muf}) this always happens at $p_T 
= 1.83$~GeV. The renormalization scale $\mu_R$, however, is 
allowed to vary with $p_T$ also below the threshold. 

In general, one would argue that the threshold values in the 
PDFs and the FFs should be the same. On the other hand, both 
PDFs and FFs are non-perturbative objects which are determined 
by fits to experimental data. Therefore, the threshold values 
$m_b^{\rm (thr)}$ used in the PDFs are also fixed by data, and 
it may turn out that different values are needed in PDFs and 
FFs. In addition, the parameter $m_b$ in Eq.~(\ref{eq:muf}) is not 
necessarily the same as either of the two threshold values in 
the FFs and PDFs. In our numerical evaluations, we shall always 
assume $m_b = 4.5$~GeV in the definition of the renormalization 
and factorization scales, Eq.~(\ref{eq:muf}), i.e.\ equal to the 
threshold value used in the FFs. With our common choice for 
$m_b$ in $\mu_i$, but differing threshold values $m_b^{\rm (thr)}$ 
in the PDFs, the turn-on of the $b$ component takes place at 
different transverse momenta. Initial-state $b$ quarks 
contribute only if $\mu_i(p_T) > m_b^{\rm (thr)}$, i.e.\ if 
the transverse momentum is large enough, while at small $p_T$ 
the $b$ component is suppressed. For CT14 and MMHT with 
$m_b^{\rm (thr)} = 4.75$~GeV, we have $\mu_i > m_b^{\rm (thr)}$ 
for $p_T \geq 3.60$~GeV, for HERAPDF2.0 with $m_b^{\rm (thr)} 
= 4.5$~GeV the transition already occurs at $p_T \geq 1.83$~GeV 
and for the recent set NNPDF3.1, where $m_b^{\rm (thr)} = 
4.92$~GeV is used, we find the threshold at $p_T \geq 
4.45$~GeV. For the CT14, MMHT and NNPDF3.1 sets, the $b$ 
component of the proton PDF vanishes for small enough $p_T$ 
values with the scale choice in Eq.~(\ref{eq:muf}). 

\begin{figure*}[t!]
\begin{center}
\raisebox{1.4mm}{
\includegraphics[width=0.48\linewidth]{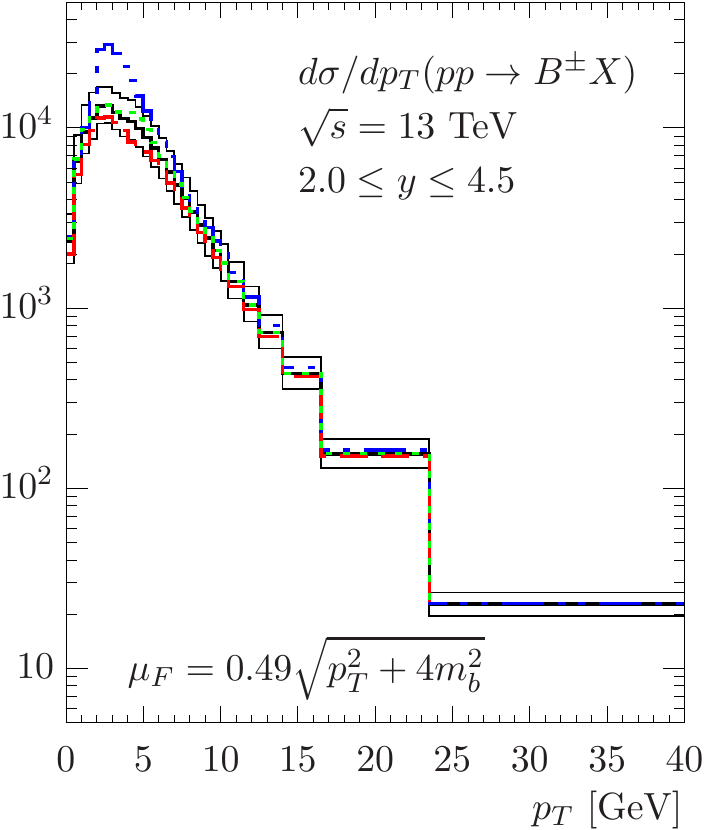}
}
\includegraphics[width=0.49\linewidth]{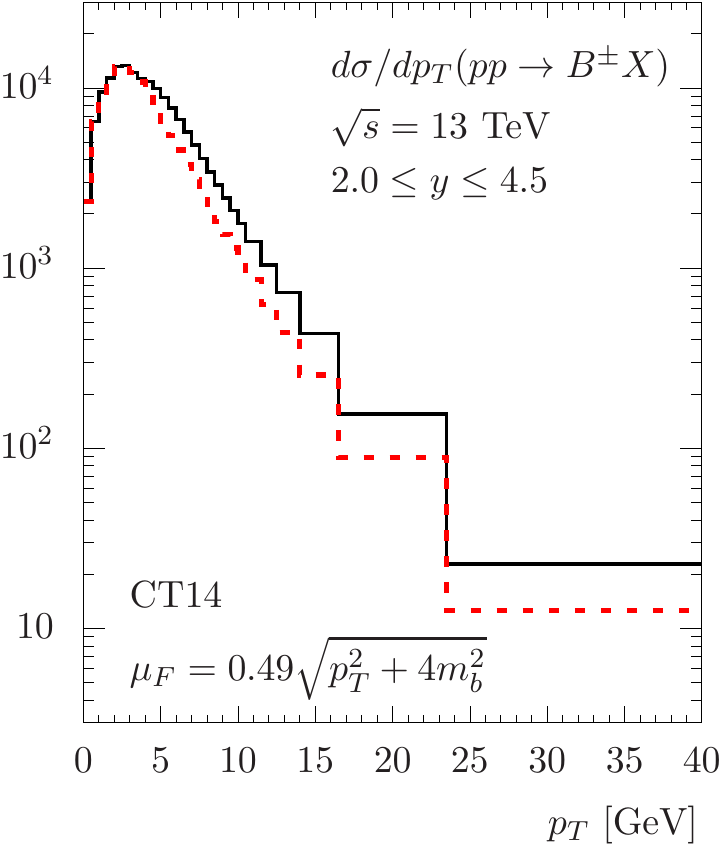}
\end{center}
\caption{
(Color online) 
The $p_T$ distribution of $B^\pm$ production at 
$\sqrt{S} = 13$~TeV integrated over rapidity $2.0 \leq y 
\leq 4.5$ with  $\mu_F = 0.49\sqrt{p_T^2 + 4 m_b^2}$, 
$m_b = 4.5$~GeV. 
Left: 
Black (full line) histogram: CT14 with $m_b^{\rm (thr)} = 
4.75$~GeV; 
green (short-dashed) histrogram: MMHT with $m_b^{\rm (thr)} 
= 4.75$~GeV; 
blue (long-short-dashed) histogram: HERAPDF2.0 with 
$m_b^{\rm (thr)} = 4.5$~GeV; 
red (long-dashed) histogram: NNPDF3.1 with $m_b^{\rm (thr)} 
= 4.92$~GeV. 
We also show the theory uncertainty band for CT14 (thin black 
line histogram). 
Right: The black (full line) histogram shows the 
complete result for the CT14 PDFs, the red (dashed-line) 
histogram shows the result without contributions due to 
$b$ quarks in the initial state.  
\label{fig:1} 
}
\end{figure*}

The influence of the different PDF sets on $d\sigma/dp_T$ 
integrated over the rapidity range $2.0 < y <4.5$ is shown 
in Fig.~\ref{fig:1}. In the left panel of this figure, the 
predictions for $d\sigma/dp_T$ obtained using the three PDF 
sets, CT14, MMHT, HERAPDF2.0 and NNPDF3.1, are compared for 
$0 \leq p_T \leq 40$~GeV as a function of $p_T$. The full 
black curve is for CT14, which agrees very well with the LHCb 
data~\cite{Aaij:2017qml}. The comparison with data will be 
discussed in more detail in the next section. The results for 
MMHT and NNPDF3.1 are very close to the CT14 curve and also agree 
with the LHCb data. All three predictions of $d\sigma/dp_T$ 
agree with each other for $p_T \geq 10$~GeV, as expected. 
Although all three predictions for $d\sigma/dp_T$ exhibit maxima 
approximately at the same $p_T$ values due to the suppression 
of the incoming $b$-quark contribution, the $p_T$ values of 
the maxima differ appreciably. For HERAPDF2.0 they are 
approximately three times larger than for CT14. 
This difference originates from the much smaller $p_T$ 
value at which the initial $b$-quark contribution of the 
proton decouples in the case of HERAPDF2.0. 

The contribution due to $b$ quarks in the initial state 
can be inferred from the results shown in the right part 
of Fig.~\ref{fig:1}. There we have plotted $d\sigma/dp_T$ 
for CT14 including all contributions (black full-line histogram, 
the same as in the left part Fig.~\ref{fig:1}) and with the 
contribution from incoming $b$ quarks subtracted (dashed red 
histogram). In the region $p_T \lsim 4$~GeV, the two histograms 
coincide, whereas for $p_T \gsim 4$~GeV one can see that the 
contribution from incoming $b$ quarks becomes more and more 
important. Similar results have been obtained for all other 
PDF sets, but the position at which the $b$-quark contribution 
sets in appears at different $p_T$ values. 

We conclude that PDF sets with a small value for the $b$-quark 
threshold should not be used in the GM-VFNS to describe $B$-meson 
production in the low-$p_T$ region. One can try to adjust the 
definition of the factorization scale $\mu_i$ to the threshold 
$m_b^{\rm (thr)}$, but this requires considerable fine-tuning 
and leads, in some cases, to additional shoulders in the $p_T$ 
distribution which worsens the agreement with the data. This 
applies to HERAPDF2.0. It would also apply to version 3.0 of 
the NNPDF PDFs \cite{Ball:2014uwa}. The NNPDF3.0 set was fitted 
to data with $m_b^{\rm (thr)} = 4.18$~GeV. The choice of this 
value was motivated by other determinations of the $b$-quark 
mass based on QCD sum rules. For NNPDF3.0, $\mu_i$ as defined 
in Eq.~(\ref{eq:muf}) is therefore always larger than the PDF 
threshold $m_b^{\rm (thr)}$, even at $p_T = 0$~GeV. We do not 
show numerical result for this PDF set because we could not 
find good agreement with the LHCb data even for very specific 
choices of the scale parameters.


\section{Cross section results and comparisons with LHCb data}

\begin{figure*}[b!]
\begin{center}
\includegraphics[width=0.49\linewidth]{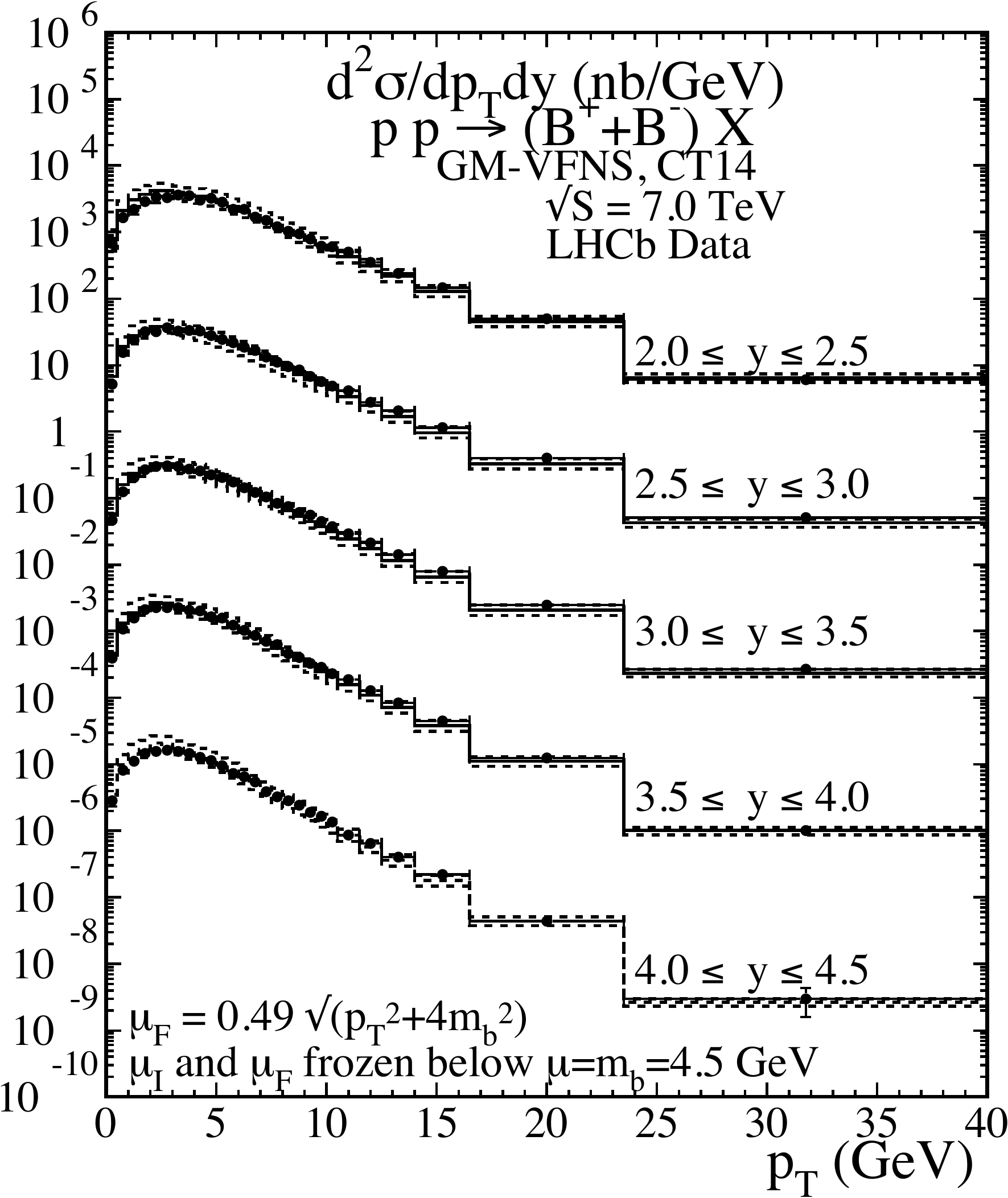}
\includegraphics[width=0.49\linewidth]{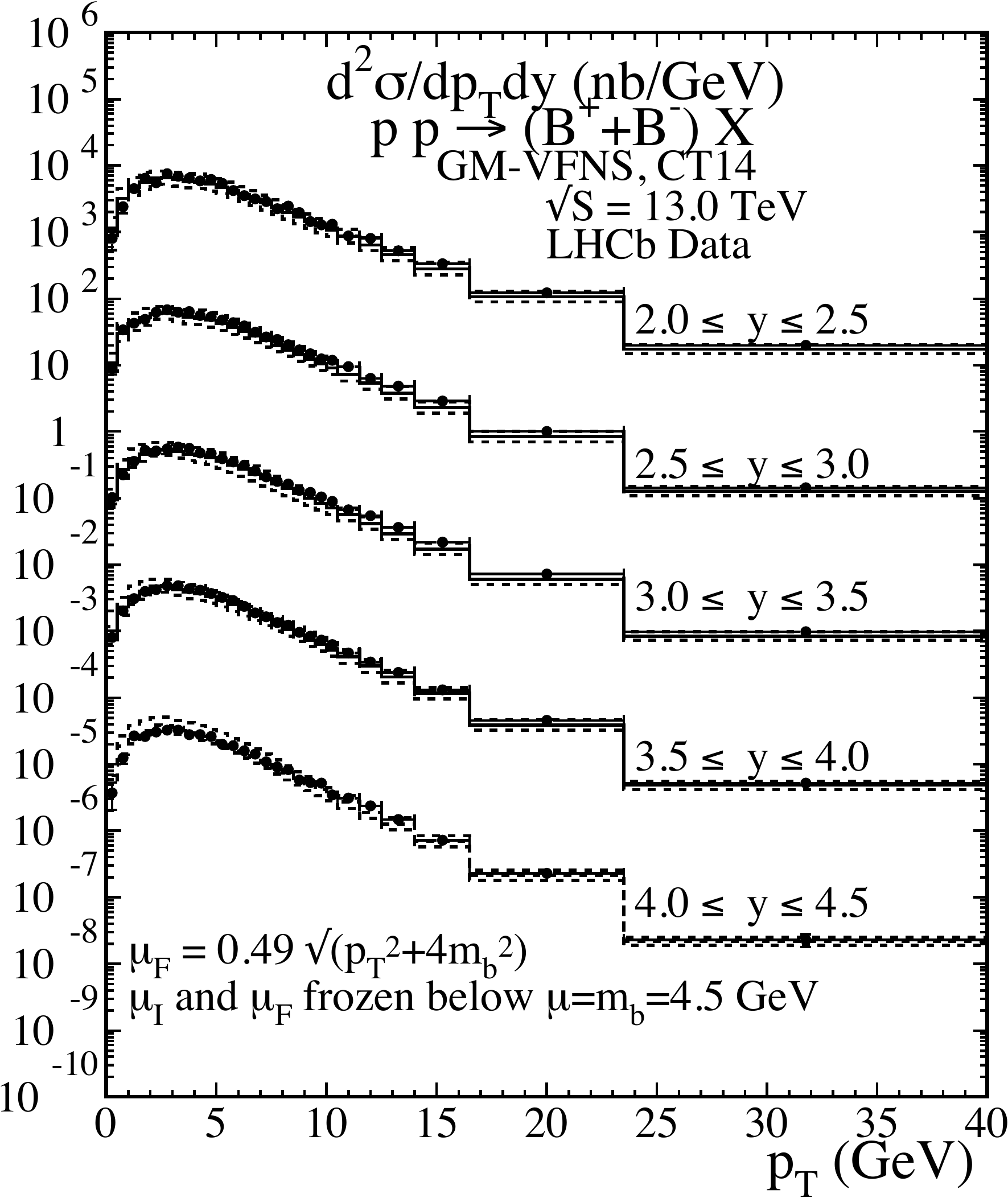}
\end{center}
\caption{
$B^{\pm}$ double-differential production cross sections at 
$\sqrt{S} = 7$~TeV (left) and $\sqrt{S} = 13$~TeV (right) 
as a function of $p_T$ and $y$. The black points represent 
the measured values from the LHCb Collaboration 
\cite{Aaij:2017qml}. The full-line histogram is the prediction 
with the default choice of scales using the CT14 PDF set. The 
dashed histograms represent the theoretical uncertainty of the 
calculated cross sections. 
\label{fig:3} 
}
\end{figure*}

\begin{figure*}[b!]
\begin{center}
\includegraphics[width=0.49\linewidth]{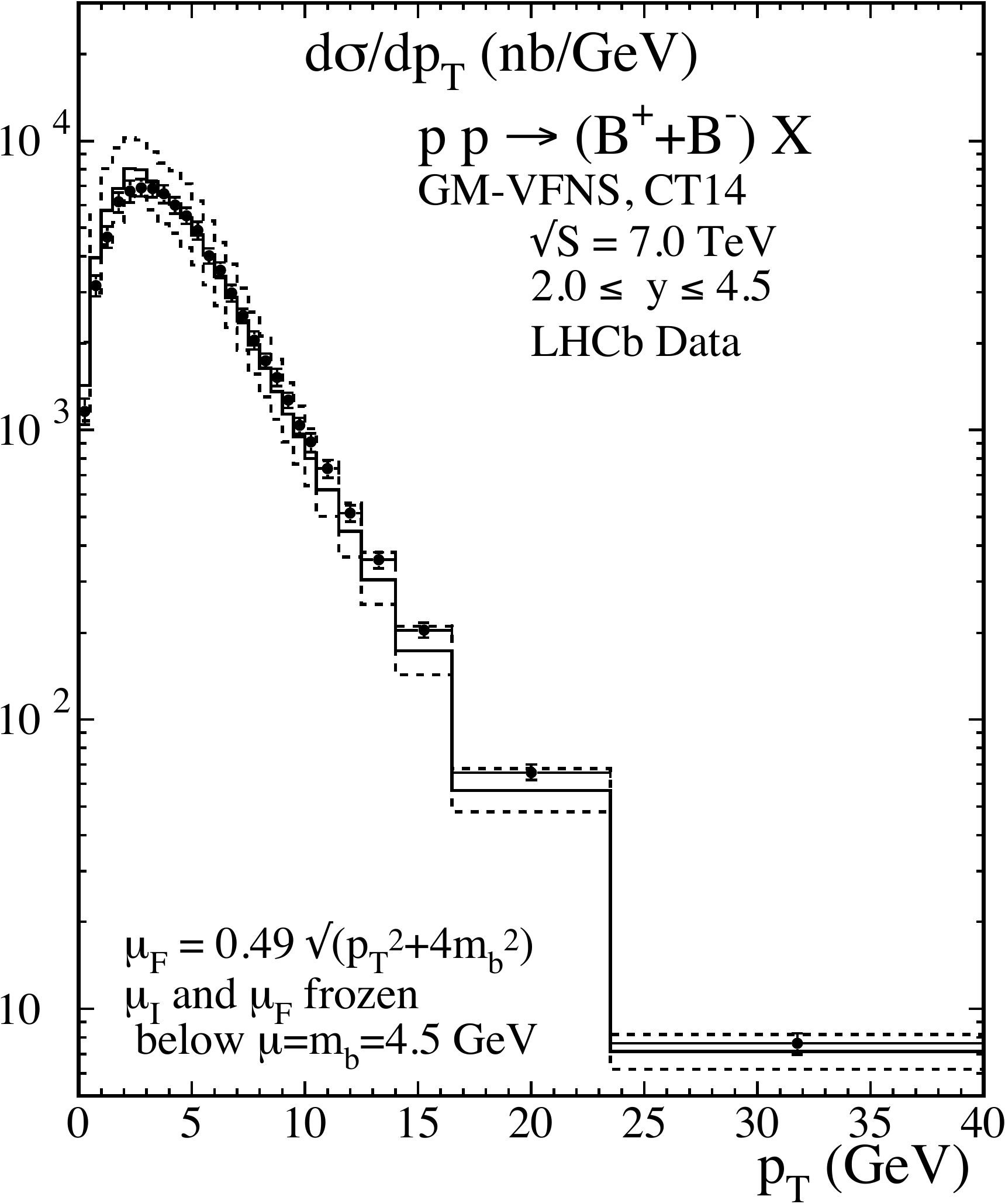}
\includegraphics[width=0.49\linewidth]{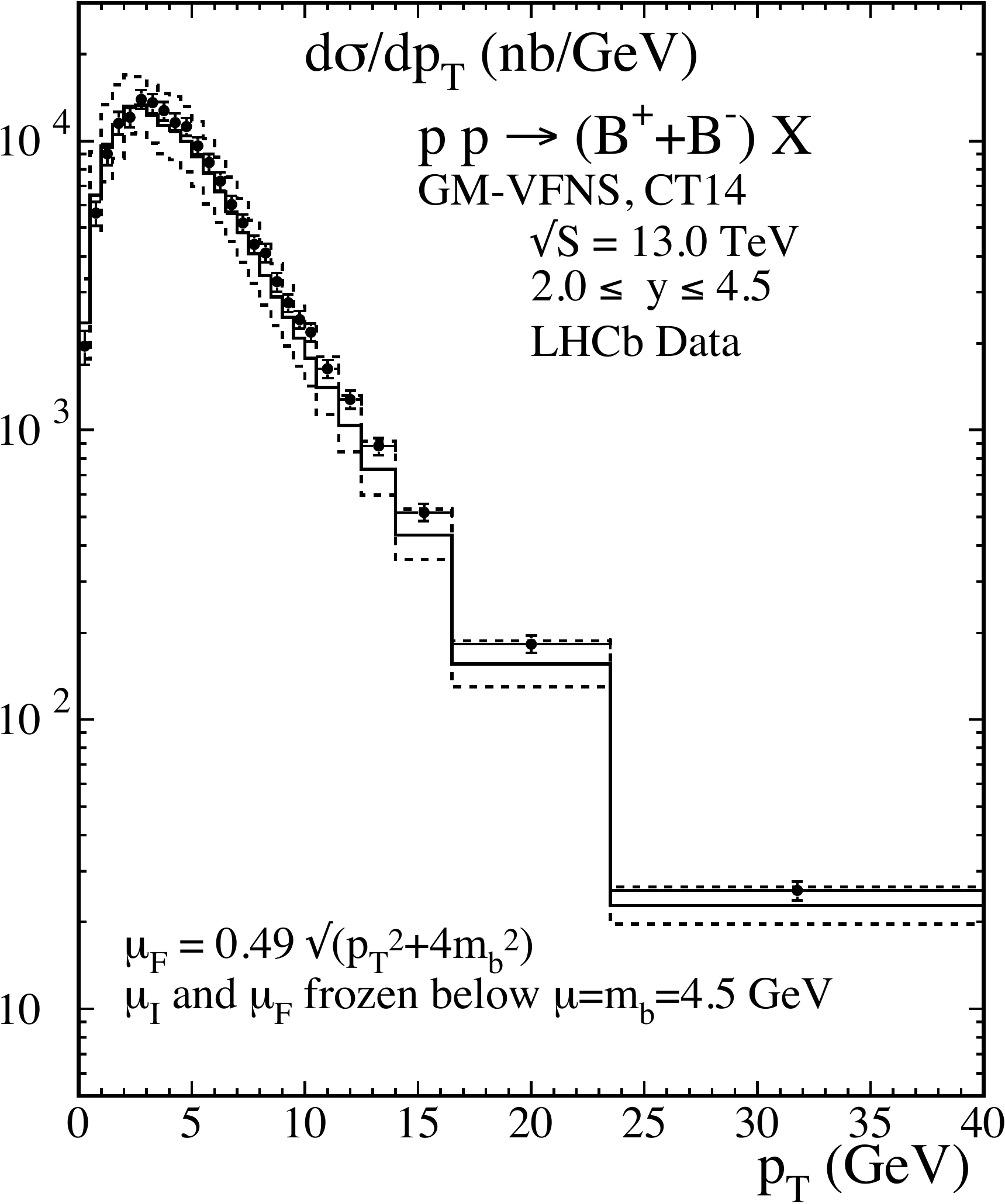}
\end{center}
\caption{
$B^{\pm}$ differential cross sections at $\sqrt{S} = 7$~TeV 
(left) and $\sqrt{S} = 13$~TeV (right) as a function of 
$p_T$, integrated over $y$ in the range $2.0 \leq y 
\leq 4.5$. The black points represent the measured values from 
the LHCb Collaboration. The full-line histogram is the 
prediction with the default choice of scales using the 
CT14 PDF set. The dashed histograms represent the theoretical 
uncertainty of the calculated cross sections. 
\label{fig:4} 
}
\end{figure*}

The LHCb Collaboration has measured the cross sections for 
$B^{\pm}$-meson production at $\sqrt{S} = 7$ and $13$~TeV in 
the transverse momentum range $0 < p_T < 40$~GeV and in five 
rapidity bins covering $2.0 < y < 4.5$. The double-differential 
cross section data, $d^2\sigma/dydp_T$, as a function of $p_T$ 
and in bins of $y$ are compared with our results for the CT14 
PDF set in Fig.~\ref{fig:3} (left plot for $\sqrt{S} = 7$~TeV, 
right plot for $\sqrt{S} = 13$~TeV). To improve readability 
of the plots, both experimental data and theoretical predictions 
are multiplied by scaling factors $10^{-2}$, $10^{-4}$, 
$10^{-6}$ and $10^{-8}$ in the $y$ bins $2.5 < y < 3.0$, 
$3.0 < y < 3.5$, $3.5 < y < 4.0$, and $4.0 < y < 4.5$, respectively. 
The full-line histogram is the default prediction with the 
factorization scale $\mu_F$ as in Eq.~(\ref{eq:muf}) 
and the renormalization scale $\mu_R = \sqrt{p_T^2 + 4m_b^2}$. 
The theoretical error is estimated by multiplying $\mu_R$ by 
factors 0.5 and 2.0, but leaving $\mu_F$ unchanged. Note that 
this implies a reduced theoretical error. The results for the 
maximal and minimal cross sections are given by the dashed 
histograms. The agreement between data and predictions is 
excellent, in particular in the small $p_T$ range, both for 
$\sqrt{S} = 7$ and 13~TeV. As shown in Ref.~\cite{Aaij:2017qml}, 
also the comparison with the FONLL prediction \cite{Cacciari:2012ny} 
exhibits a similarly good agreement between data and predictions.

In Ref.~\cite{Aaij:2017qml}, also the corresponding single 
differential cross sections $d\sigma/dp_T$ obtained from the 
measured double-differential cross sections by summing over 
the five rapidity bins are given. These data are shown in 
Fig.~\ref{fig:4} and compared with our predictions for 
$\sqrt{S} = 7$ and 13~TeV. The agreement between data and 
predictions is equally good as above for the double-differential 
cross sections shown in Fig.~\ref{fig:3}. 

\begin{figure*}[b!]
\begin{center}
\includegraphics[width=0.49\linewidth]{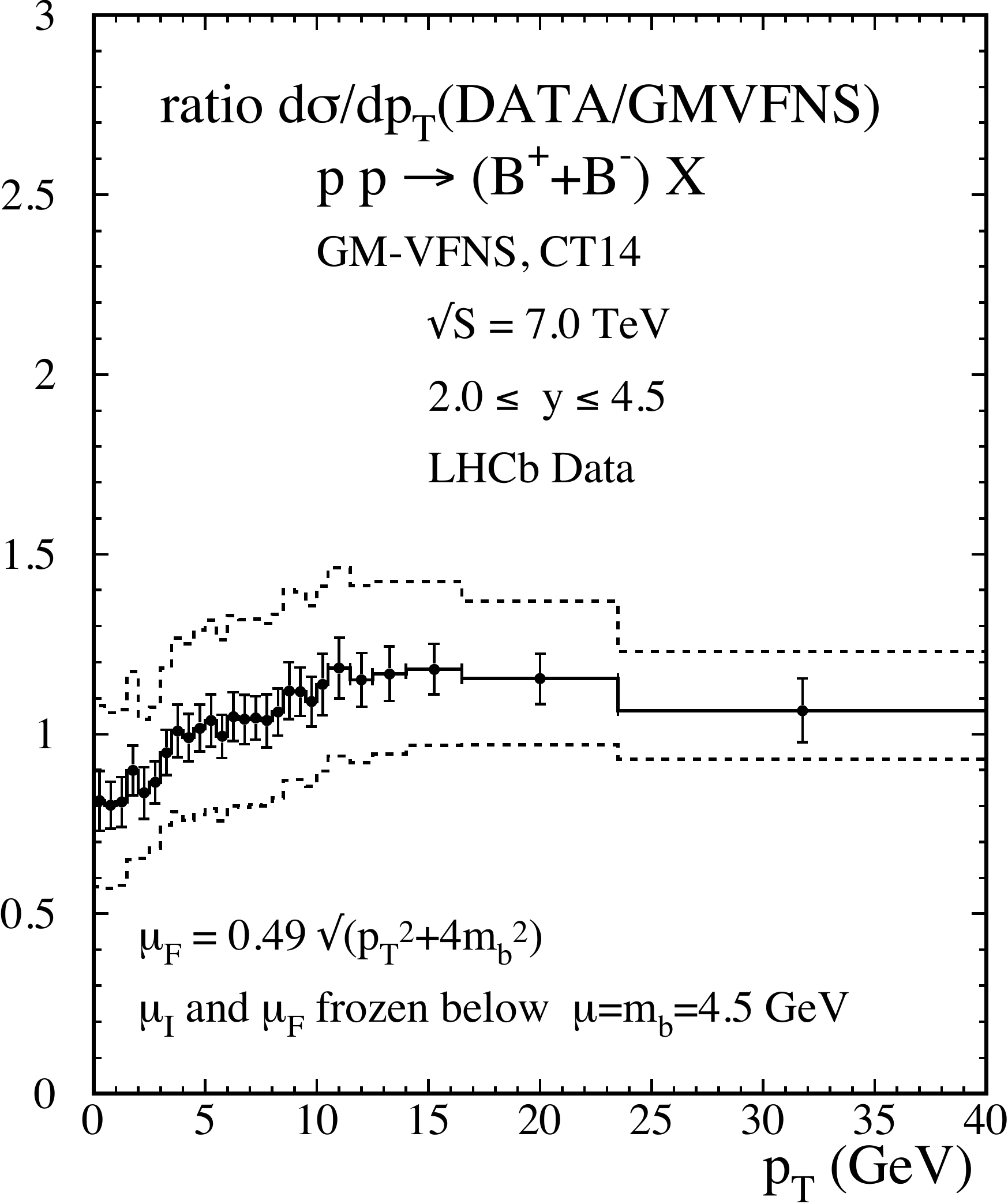}
\includegraphics[width=0.49\linewidth]{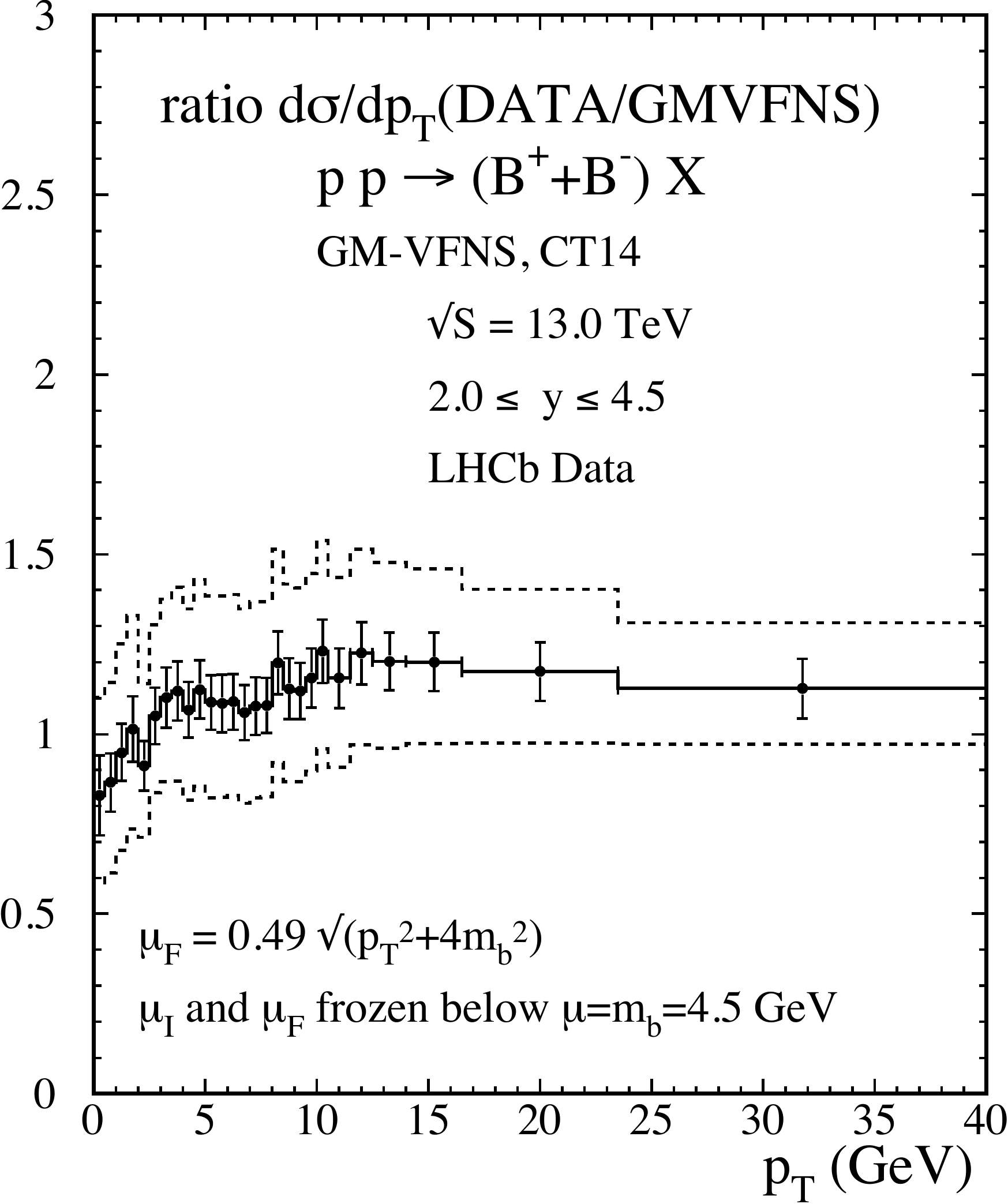}
\end{center}
\caption{
Ratio of the measured over predicted $B^{\pm}$ $p_T$ 
distributions at $\sqrt{S} = 7$~TeV (left) and 13~TeV (right), 
integrated over $y$ in the range $2.0 \leq y \leq 4.5$. 
The black points connected by the full-line histogram represent 
the ratio calculated with the default scales. The error bars 
represent the experimental uncertainties and the dashed 
histograms show the theoretical uncertainties obtained by 
varying the renormalization scale. The PDF set CT14 was used. 
\label{fig:5} 
}
\end{figure*}

The ratios of data over predictions are shown in Fig.~\ref{fig:5}. 
From this plot, one can see more clearly the quality of the 
agreement, which is better than 20 $\%$ and well inside the 
theoretical uncertainties. The majority of the data points 
agree with the predictions also within the smaller experimental 
errors; only in the intermediate range $10 \lsim p_T \lsim 
20$~GeV, the data tend to lie slightly above the predictions. 

\begin{figure*}[b!]
\begin{center}
\includegraphics[width=0.49\linewidth]{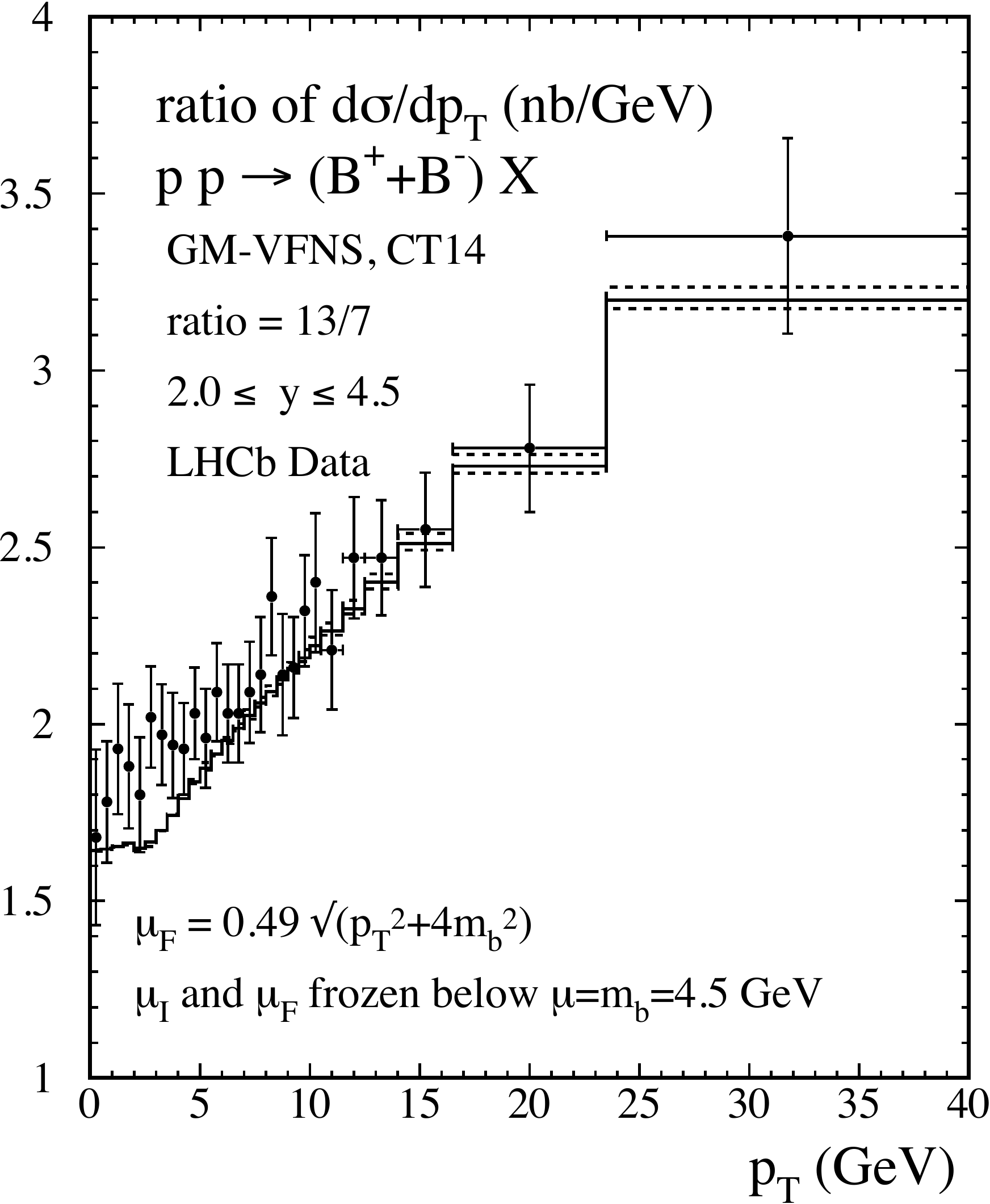}
\end{center}
\caption{
Ratio of the $B^{\pm}$ $p_T$ distribution at $\sqrt{S} = 13$~TeV 
to that at $\sqrt{S} = 7$~TeV, integrated over $y$ in the 
range $2.0 \leq y \leq 4.5$. The black points represent the 
measured values from the LHCb Collaboration. The full-line 
histogram is the prediction with the default choice of scales 
using the PDF set CT14. The dashed histograms represent the 
theoretical uncertainties. 
\label{fig:6} 
}
\end{figure*}

Fig.~\ref{fig:6} shows the ratio of $d\sigma/dp_T$ at the 
two center-of-mass energies, $R_{13/7} = d\sigma(\sqrt{S} 
= 13~\mbox{TeV}) / d\sigma(\sqrt{S} = 7~\mbox{TeV})$. The 
data are taken from Ref.~\cite{Aaij:2017qml} and compared with 
our predictions obtained as the ratios of the cross sections 
shown in Fig.~\ref{fig:4}. The comparison shows good agreement. 
The theoretical uncertainty of the ratio is quite small, 
since the scale variation is performed in the numerator 
and denominator of the ratio in the same way. 

\begin{figure*}[b!]
\begin{center}
\includegraphics[width=0.48\linewidth]{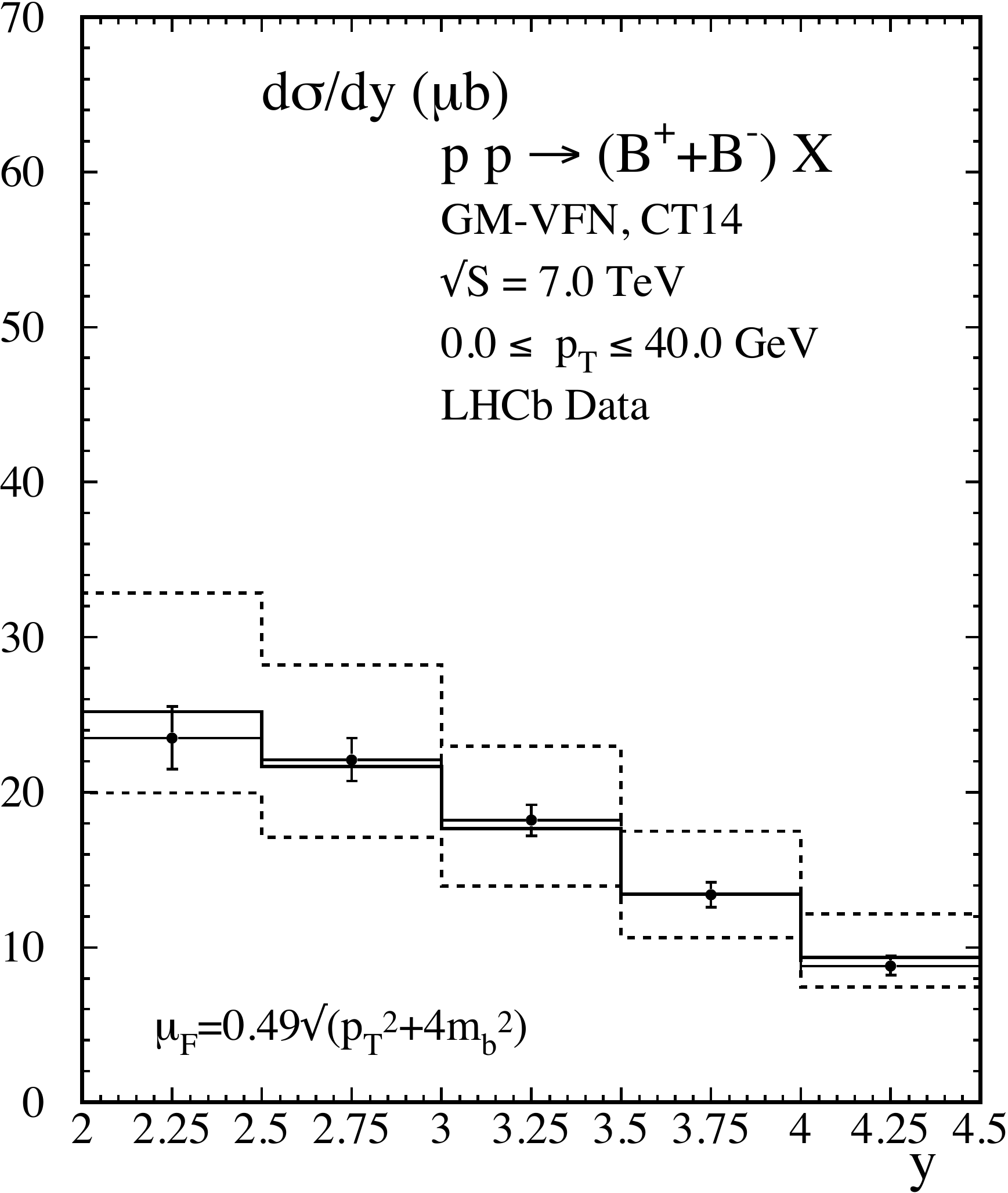}
\includegraphics[width=0.49\linewidth]{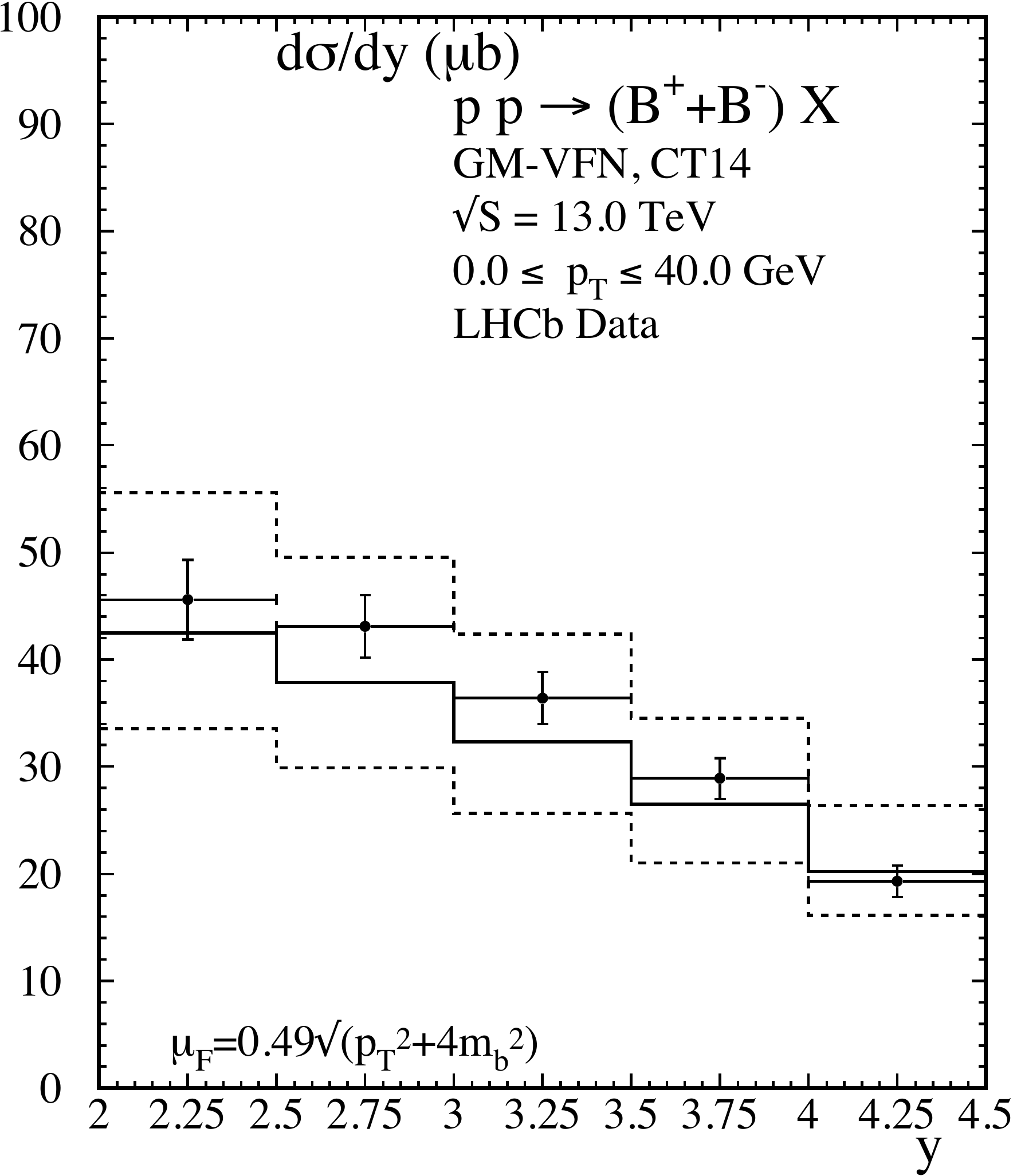}
\end{center}
\caption{
$y$ distribution of $B^{\pm}$ production at $\sqrt{S} = 7$~TeV 
(left) and 13~TeV (right), integrated over $p_T$ 
in the range $0 \leq p_T \leq 40$~GeV. The black points represent 
the measured values from the LHCb Collaboration. The full-line 
histogram is the prediction with the default choice of scales 
using the CT14 PDF set. The dashed histograms represent the 
theoretical uncertainties. 
\label{fig:7} 
}
\end{figure*}

\begin{figure*}[b!]
\begin{center}
\includegraphics[width=0.49\linewidth]{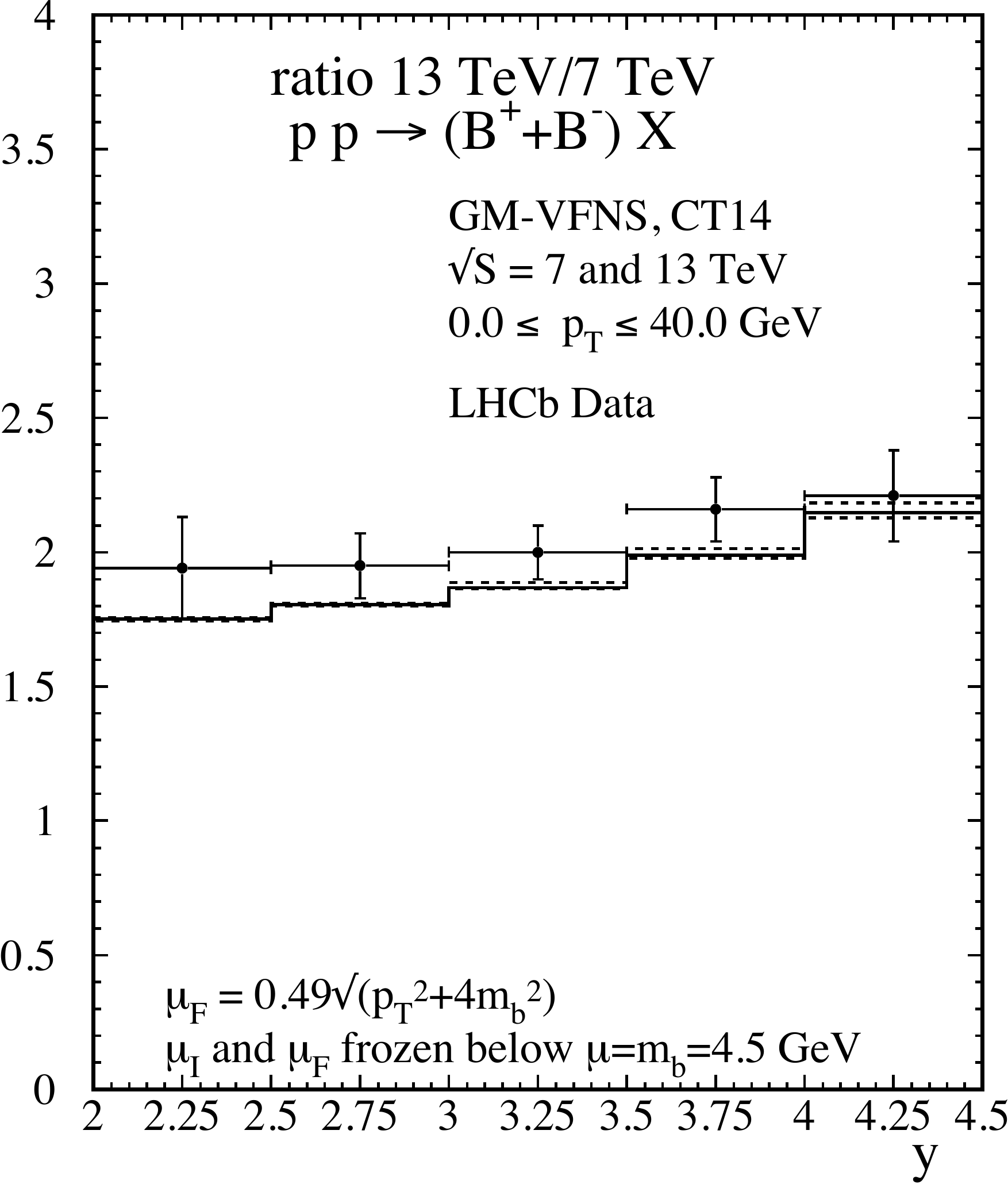}
\end{center}
\caption{
Ratio $R_{13/7}$ of the $B^{\pm}$ $y$ distributions at 
$\sqrt{S} = 13$~TeV and 7~TeV, integrated over $p_T$ in 
the range $0 \leq p_T \leq 40$~GeV. The black points 
represent the measured values from the LHCb Collaboration. 
The full-line histogram is the prediction with the default choice 
of scales using the CT14 PDF set. The dashed-line histograms 
represent the theoretical uncertainties. 
\label{fig:8} 
}
\end{figure*}

Finally we present results for the rapidity distributions. 
In Fig.~\ref{fig:7}, we show the cross section $d\sigma/dy$, 
integrated over the range $0 \leq p_T \leq 40$~GeV, for 
$\sqrt{S} = 7$ and 13~TeV and for the five $y$ bins in the 
range $2.0 < y < 4.5$. As expected from the agreement seen 
for the double-differential cross sections, our predictions 
of the $y$ dependence also agrees well with the data of the 
LHCb Collaboration~\cite{Aaij:2017qml}. The ratio $R_{13/7}$ 
as a function of $y$ is shown in Fig.~\ref{fig:8}. We observe 
once more good agreement between the data from 
Ref.~\cite{Aaij:2017qml} and the GM-VFNS predictions. 
 
\section{Conclusions}

We have performed a detailed analysis of $B$-meson production 
at NLO in the perturbative-QCD framework of the general-mass 
variable-flavour-number scheme. The recent high-quality data 
from the LHCb Collaboration can be described in this framework 
over a large range of transverse momenta, down to $p_T = 0$. 
Both the $p_T$ and $y$ distributions, as well as the cross 
section ratios taken at different center-of-mass energies, 
agree well with data. 

We found it to be crucial in this comparison that the 
transition at small $p_T$ values to the fixed-flavour-number 
scheme is implemented in a proper way. With a judicious choice 
of the factorization scale it is possible to turn off contributions 
from initial-state $b$ quarks in the hard-scattering processes. 
This works, however, only when the heavy-quark threshold in 
the PDF parametrization matches the corresponding threshold 
chosen in the FFs. At present, all modern FF sets compatible 
with the GM-VFNS \cite{Kniehl:2008zza,Salajegheh:2019ach} have 
been obtained with the same fixed value of $m_b = 4.5$~GeV. This 
limits the possible choice of PDFs which lead to a consistent 
framework. Future improvements of PDF fits will require for 
more and more precise data to be included, and it is expected 
that inclusive heavy-quark production will play an important 
role for that. It may become important in this challenge to 
consider both PDFs and FFs on the same footing within a common 
fit to data. 
\\

\section*{Acknowledgment}

B.~A.~K.\ was supported in part by the German Federal Ministry for Education and Research (BMBF) through Grant No.\ 05H18GUCC1.
We thank M.~V.~Garzelli for useful discussions. 

\clearpage


\end{document}